\documentclass[11pt]{article}
\usepackage{a4wide}                       
\usepackage[dvips]{graphicx}
\usepackage{epsf}                        
\usepackage{latexsym}                    
\usepackage{amsfonts}                    
\usepackage{amssymb}                     

\newcommand{\sect}[1]{\setcounter{equation}{0}\section{#1}}

\def\be{\begin{equation}}
\def\ee{\end{equation}}
\def\bea{\begin{eqnarray}}
\def\eea{\end{eqnarray}}
\def\nn{\nonumber \\}
\def\nnw{\nonumber \\ [.2cm]}

\def\part{\partial}
\def\tfrac#1#2{{\textstyle{\frac{#1}{#2}}}}
\def\half{\tfrac{1}{2}}
\def\x{\times}

\def\Str{\mbox{STr}}
\def\STr{\mbox{STr}}

\def\incl{\mbox{i}}

\def\Ortin{Ort{\'\i}n}

\def\R{\ensuremath{\mathbb{R}}}

\def\mn{{\mu\nu}}

\def\mnrl{{\mu\nu\rho\lambda}}

\def\makeatletter{\catcode`\@=11}
\makeatletter
\def\mathbox#1{\hbox{$\m@th#1$}}%
\def\math@ccstyles#1#2#3#4#5#6#7{{\leavevmode
      \setbox0\mathbox{#6#7}%
      \setbox2\mathbox{#4#5}%
      \dimen@ #3%
      \baselineskip\z@\lineskiplimit#1\lineskip\z@
      \vbox{\ialign{##\crcr
             \hfil \kern #2\box2 \hfil\crcr
             \noalign{\kern\dimen@}%
             \hfil\box0\hfil\crcr}}}}
\def\mathaccstyles{\math@ccstyles\maxdimen}
\def\maththroughstyles{\math@ccstyles{-\maxdimen}}
\def\unity%
 {\maththroughstyles{.45\ht0}\z@\displaystyle {\mathchar"006C}\displaystyle 1}
\begin{document}
\rightline{UG-FT-203/06}
\rightline{CAFPE-73/06}
\rightline{FFUOV-06/09}
\rightline{FTUAM-06/07}
\rightline{hep-th/0606264}
\rightline{October 2006}
\begin{center}
\huge \bf The baryon vertex with magnetic flux 
\end{center}

\vspace{1truecm}

\centerline{
    {\large \bf Bert Janssen${}^{a,}$}\footnote{E-mail address: 
                                  {\tt bjanssen@ugr.es} },
    {\large \bf Yolanda Lozano${}^{b,}$}\footnote{E-mail address:
                                  {\tt yolanda@string1.ciencias.uniovi.es}}
    {\bf and} 
    {\large \bf Diego Rodr\'{\i}guez-G\'omez${}^{c,}$}\footnote{E-mail address: 
                                  {\tt Diego.Rodriguez.Gomez@cern.ch}}
                                                            }
\vspace{.4cm}
\centerline{{\it  ${}^a$Departamento de F\'{\i}sica Te\'orica y del Cosmos and}}
\centerline{{\it Centro Andaluz de F\'{\i}sica de Part\'{\i}culas Elementales}}
\centerline{{\it Universidad de Granada, 18071 Granada, Spain}} 

\vspace{.4cm}
\centerline{{\it ${}^b$Departamento de F{\'\i}sica,  Universidad de Oviedo,}}
\centerline{{\it Avda.~Calvo Sotelo 18, 33007 Oviedo, Spain}}

\vspace{.4cm}
\centerline{{\it ${}^c$Departamento de F{\'\i}sica Te\'orica C-XI,}}  
\centerline{{\it Universidad Aut\'onoma de Madrid,}}
\centerline{{\it Cantoblanco, 28049 Madrid, Spain}}

\vspace{1truecm}

\centerline{\bf ABSTRACT}
\vspace{.5truecm}

\noindent
In this letter we generalise the baryon vertex configuration of AdS/CFT 
by adding a suitable instantonic magnetic field on its worldvolume,
dissolving D-string charge.
A careful analysis of the configuration shows that there is an upper
bound on the number of dissolved strings. This could be a
manifestation of the stringy exclusion principle. We provide a
microscopical description of this configuration in terms of a dielectric effect
for the dissolved strings.

\newpage
\sect{Introduction}

The $AdS$/CFT duality \cite{Maldacena} relates gravity theories in $AdS$
spaces with certain conformal field theories. In particular, it states that
IIB string theory on $AdS_5\times S^5$ is dual to $\mathcal{N}=4$ SYM in four
dimensions. In the context of the dual field theory, an extremely interesting
question is whether the theory is confining or not. In order to study this, the
appropriate quantity to look at is the Wilson loop, whose VEV gives the
quark-antiquark ($q\bar{q}$) potential. In the case at hand, $\mathcal{N}=4$ SYM has no
dynamical quarks. However, one can introduce static quarks and compute the
appropriate Wilson loop to obtain the $q\bar{q}$ potential. The Wilson
loop can be computed both at weak 't Hooft coupling, directly in field theory
\cite{ESZ}, and at strong 't Hooft coupling, in the gravity side of the
correspondence \cite{Maldacena2,RY}.
In both regimes the $q\bar{q}$ potential goes
like $1/d$, where $d$ is the distance between the quark and
antiquark. This particular coulombian shape, which
exhibits no confinement, is due to the conformal invariance of the theory.

As shown in \cite{Maldacena, RY}, F-strings ending on the D3-brane
and going all the way to the $AdS$ boundary are seen in the dual theory as
external quarks or antiquarks, depending on the string's orientation. It is clear then
that one can form a $q\bar{q}$ state with a single string coming from the boundary
to the D3-brane and then going back to the boundary. Once the
coupling to the $SU(N)$ theory is taken into account, the string ends up being  U-shaped,
with the apex at a distance $u_0$ from the stack of D3-branes. This configuration
is seen as a $q\bar{q}$ pair on the SYM side, whose energy is computed by
means of a rectangular Wilson loop. On the other hand, on the gravity side the energy is 
computed by minimising the worldsheet area of the string ending on the loop.

{}From the Wilson loop one can only extract the $q\bar{q}$ potential. However,
 one would expect that there should be a mechanism to form bound states of $N$ 
non-dynamical quarks. In \cite{Witten} precisely this  
question was asked, namely whether it would also be possible to construct in this set-up 
a baryon configuration. Roughly speaking, a baryon is a colourless bound 
state of quarks with finite energy. In the case at hand, where there are no dynamical 
quarks, it turns out to be possible to construct such a bound state with static external quarks. 
In \cite{Witten} the gravitational dual of this bound state of quarks was found in terms of a 
D5-brane wrapping the $S^5$ part of the spacetime geometry.
On this D5-brane there are $N$ F-strings attached, 
stretching from the D5-brane to the boundary of $AdS_5$. The endpoints of the $N$ F-strings 
are then regarded on the dual SYM side as a bound state of $N$ 
quarks, in other words, as a baryon.\footnote{Since the quarks are non-dynamical, 
  this represents a mechanism to form the baryon, and is referred to as the 
  \textit{baryon vertex}.} 
Indeed it can be shown \cite{Witten} that the associated wave function satisfies the required 
symmetry properties. 

In this letter we generalise this baryon vertex configuration by adding a 
new quantum number. The key point is to realise that $S^5$ can be seen as an 
$S^1$ bundle over $CP^2$. The $S^1$ fibre is a non-trivial 
$U(1)$ gauge bundle on the $CP^2$ base, and this allows to switch on a magnetic 
BI field  on the worldvolume of the D5-brane, proportional to the
curvature tensor of the
fibre connection. 
As we will see, the effect of this field is to dissolve D1-branes 
wound around the $S^1$  direction on the D5-brane. 

The interest of this generalised baryon vertex is twofold. 
On one hand, the analysis of the equations of motion reveals
that there is a bound on the number of 
D-strings that can be dissolved in the D5-brane.
This is an interesting phenomenon, which could be related to
the stringy exclusion principle \cite{MaldaStrom}. Indeed, by dissolving
D-strings in the configuration we are inducing a non-zero winding
charge along a cycle of the $S^5$, and these
winding charges appear in the dual 
field theory as non-zero charges  under certain $U(1)$ subgroups of the
$SO(6)$ R-symmetry \cite{BHK}, which are bounded due to conformal invariance.
A complete analysis in the field theory context is however beyond the scope of this
letter.

On the other hand,
the fact that we have dissolved D-strings on the worldvolume of the D5-brane hints 
at the existence of an alternative microscopical description in terms of non-Abelian
D-strings polarising due to a dielectric effect \cite{Myers}. We give 
such a microscopical description in terms of D1-branes expanding into
a fuzzy spherical D5-brane using 
the action of \cite{Myers}. We also consider the S-dual of
the baryon vertex with magnetic flux, which consists on a spherical NS5-brane
with dissolved F-strings, and with N D1-branes attached to it. We show
that this
configuration can also be described microscopically in terms of F-strings
expanding into a fuzzy spherical NS5-brane by dielectric effect.

This letter is organised as follows. In section 2 we present the D5-brane 
description of the generalised baryon vertex. We start in subsection 2.1 by 
revisiting the construction of the original baryon vertex as given by 
\cite{Witten}, and then generalise this construction in subsection 2.2 to include 
a magnetic BI vector on the worldvolume. 
In subsection 2.3 we study the dynamics of this configuration
and show that there is a bound on the number of dissolved 
strings. Section 3 is devoted to the microscopic description of the generalised
baryon vertex in terms of non-Abelian D-strings. 
In subsection 3.1 we calculate the energy
of the configuration of multiple coinciding D-strings polarising into a
fuzzy spherical D5-brane. In subsection 3.2 we show how the $N$
fundamental strings that connect the (dielectric) D5-brane to the
gauge theory on the boundary arise in the microscopical set-up. Subsection 3.3
contains the description of the S-dual of the baryon
vertex with magnetic flux in terms of fundamental strings expanding
into a fuzzy spherical NS5-brane. The action describing coinciding
fundamental strings is constructed from the action for coinciding Type IIA
gravitational waves of \cite{JL2} using T-duality.
In the conclusions we review the main points of our construction.

\sect{The baryon vertex with magnetic flux}

\subsection{The baryon vertex revisited}

We start by reviewing the major points in the construction of the baryon
vertex, as given in \cite{Witten}. Consider a probe D5-brane wrapped on the
5-sphere and static in a fixed point in $AdS$. In the $AdS_5 \x S^5$ background 
there is no 6-form R-R potential to which the probe brane can couple, however the 
presence of the 4-form R-R field in the Chern-Simons action induces a coupling to the
BI field strength $F=dA$ of the form 
\be
S_{CS}=-T_5\int_{\R\times S^5} P[C^{(4)}]\wedge F.
\label{C^4F}
\ee
In our specific setting, the only non-zero contribution is that of
the coupling of the magnetic part of the R-R form to the electric component of
$F$. Integrating by parts, we find that this term can be rewritten as
\be
S_{CS}= \ T_5\int_{\R\times S^5} P[G^{(5)}]\wedge A,
\label{G^5A}
\ee
where $G^{(5)} = dC^{(4)}$ is the R-R 5-form field strength. In our
particular background, we have that  $G^{(5)} =4L^4 \sqrt{g_{S^5}}$, such that
$\int_{S^5}G^{(5)}=4\pi^2 N$ (in units where $2\pi l_s^2=1$), with $N$ the number 
of D3-branes that build up the
background. If we therefore take as an Ansatz for the BI vector
\be
A = A_t (t) dt,
\label{AnsatzA}
\ee
it is clear that the coupling (\ref{G^5A}) factorises as
\begin{equation}
\label{SCSD5}
S_{CS}=T_5\int_{S^5}G^{(5)}\int dt A_t\ =\ N T_1 \int dt A_t,
\end{equation}
where we have taken into account that the tension of the D5-brane and
the tension of a string are related
by $4\pi^2 T_5=T_1$. Therefore,
one can interpret that the coupling (\ref{C^4F}) is inducing
$N$ units of BI electric charge on the D5-brane, such that the total action for the
wrapped D5-brane can be written as 
\begin{equation}
S = S_{DBI} + N T_1\int dt A_t.
\label{SD5}
\end{equation}
However we have to check whether the Ansatz (\ref{AnsatzA}) is consistent with the 
equations of motion of the D5-brane system (\ref{SD5}). As (\ref{AnsatzA}) implies 
that $F=0$, it is clear that the equation of motion of $A$ is given by
\begin{equation}
0\equiv \frac{\partial \mathcal{L}}{\partial A_t}=N T_1.
\end{equation}
In other words, the equations of motion imply that the Ansatz
(\ref{AnsatzA}) is only compatible with the action (\ref{SD5}) if the total BI
electric charge on the D5-brane is zero, as it is wrapped on a compact manifold. However,
there is a consistent way of inducing a non-zero BI electric charge in the worldvolume of
the D5-brane, by cancelling this charge with the charge induced by the endpoints of $N$ 
open fundamental strings (with appropriate orientation) stretching between the D5-brane and
the boundary of the $AdS$ space. The action (\ref{SD5}) is therefore not describing the
entire system, but only the D5-brane part. In order to describe the full
dynamics one has to add the action for the open strings, consisting of $N$
copies of the Nambu-Goto action $S_{F1}$, and a boundary term contribution
$T_1\int A_t dt$ from the endpoints: 
\be
S_{\rm total} = S_{DBI}   +  N T_1 \int dt A_t \ \ + \ \ N S_{F1}  -  N T_1\int
dt A_t\, .
\ee
Note that the contribution from the open string endpoints cancels
exactly the Chern-Simons term in the D5-brane action, such that the total
system is described by \cite{BISY}
\be
S_{\rm total} = S_{DBI}\ \ + \ \ N S_{F1}.
\label{S=S5+S1}
\ee
The configuration that we have just described
is the so-called baryon vertex. Since the $N$
F-strings, stretching from the D5-brane all the way to the $AdS$ boundary, 
have the same orientation, the dual configuration on the CFT side corresponds
to the bound state of $N$ (anti)quarks, which is gauge invariant and
antisymmetric under the interchange of any two quarks \cite{Witten}.

\subsection{Adding magnetic flux to the baryon vertex}

It is well known that $S^5$ can be regarded as a $U(1)$ fibre over
$CP^2$ with a non-trivial fibre connection.
{}From the 
$CP^2$ point of view, the $U(1)$ connection, $B$, can be seen as a non-trivial gauge bundle 
inducing a non-zero instanton number \cite{Trautman,HP}.
In view of this, it seems
natural to consider a generalisation of the baryon vertex in which
magnetic components of the BI field strength are switched on, which
are proportional to $dB$.

In the $S^5$ fibre coordinates the $AdS_5\times S^5$ background reads
\bea
&& ds^2\ =\ \frac{u^2}{L^2}\eta_{ab} dx^adx^b\  + \ \frac{L^2}{u^2}du^2 \ 
      + \ L^2 \Bigl( (d\chi-B)^2+ds_{CP^2}^2\Bigr),\nnw
&&C_{abcd} = L^{-4}u^4 \epsilon_{abcd}\ , \hspace{2cm}
C_{\varphi_2\varphi_3 \varphi_4\chi} = \frac{1}{8} L^4 \sin^4 \varphi_1 \sin \varphi_2,
\label{background}
\eea
where $ds_{CP^2}^2$ stands for the Fubini-Study metric on $CP^2$,
$\chi$ is taken along the $U(1)$ fibre and $B$ is the connection of the fibre
bundle. Explicitly \cite{Pope}
\bea
&& B=-\frac{1}{2}\sin^2\varphi_1(d\varphi_4+\cos\varphi_2d\varphi_3), \nnw
&& ds_{CP^2}^2 = d\varphi_1^2  + \frac{1}{4}\sin^2\varphi_1
         \Big( d\varphi_2^2 + \sin^2\varphi_2d\varphi_3^2
                         + \cos^2\varphi_1 \big(d\varphi_4 + \cos\varphi_2d\varphi_3\big)^2\Big).
\label{CP2}
\eea
The fibre connection $B$ satisfies the following
properties \cite{Trautman}
\be
dB={}^\star(dB), \hspace{2cm}
\int_{CP^2} dB \wedge dB =4\pi^2\, , 
\label{propertiesB}
\ee 
where the Hodge star is taken with respect to the metric (\ref{CP2})
on $CP^2$.

In this system of coordinates the baryon vertex consists on the D5-brane wrapped around
the $S^5$ and the fundamental strings laying in the $u$-direction of 
$AdS_5$ \cite{RY}. 
As mentioned above, besides the electric components of the BI field strength,
representing the charges induced by the F-strings ending on the D5-brane, 
one could think of turning on also magnetic components. Due to the fact that 
$CP^2$ allows instanton solutions, it is natural to take the magnetic components 
living in the $CP^2$ and proportional to the curvature tensor of the $U(1)$ fibre connection $B$,
\be
F = 2n \, dB\, .
\label{F=NdB}
\ee
With this Ansatz $F$ satisfies the same properties
(\ref{propertiesB}) as the fibre connection $dB$,  namely it is selfdual and
\begin{equation}
\int_{CP^2}F\wedge F=  8\pi^2 n^2\, .
\label{FF=n}
\end{equation}
This integral is non-zero because it is the product of two integrals $\oint F$ over 
non-trivial two-cycles in $CP^2$. Since $\oint F=2\pi n$ due to the Dirac quantization condition, 
$n$ represents the winding number of D3-branes wrapped around each of the 
two-cycles. Note that the winding number must be the same on each cycle in order to
preserve the selfduality condition. Moreover, if we want that some of
the supersymmetries of the D5-brane, if any, are preserved,
the two D3-branes must be wrapped with the
same orientation.

With this
choice for the BI field strength it is clear that there are no other couplings
in the Chern-Simons action besides the ones we already considered in (\ref{C^4F}). The 
Born-Infeld action however is given by
\begin{equation}
S_{DBI}\ =\ -T_5\int d^6 \xi \ \frac{u}{L}\ 
                      \sqrt{\det \Bigl( g_{\alpha\beta} + F_{\alpha\beta}\Bigr)},
\label{DBI}
\end{equation}
where the coordinates $x^\alpha$ indicate the angles on the $S^5$. Since $F$ is 
selfdual, the determinant under the square root is a perfect square, yielding
\bea
S_{DBI} &=& -T_5  \int d^6 \xi \  u\ \sqrt{g_{S^5}} \ 
             \Bigl( L^4 + 2 F_{\alpha\beta}F^{\alpha\beta}\Bigr).
\label{DBI2}
\eea

The Ansatz (\ref{F=NdB}) is consistent with the
action (\ref{DBI2}), as is reflected in the fact that the equations of motion for the
magnetic components of $F$ are given by $dF= 0$, which is indeed satisfied by
(\ref{F=NdB}). Finally, substituting the expression for $F$ in the action and
integrating over the $S^5$ directions we obtain the following expression for the
energy of the spherical D5-brane:  
\begin{equation}
E_{D5}\ = \ 8\pi^3  T_5\ u \ \Big(n^2 +\frac{L^4}{8}\Big).
\label{ED5}
\end{equation}
Note that this energy consists of two parts: one contribution from the tension of the 
5-brane wrapped around the five-sphere and one from the magnetic flux of the
BI vector. 

While the electric components of $F$ induce $N$ units of BI charge on the D5-brane worldvolume
through the coupling (\ref{C^4F}), the magnetic components induce a non-zero instanton number $n^2$, 
due to (\ref{FF=n}). In particular, the Chern-Simons coupling
\begin{equation}
S_{CS}=\frac{1}{2}T_5\int_{\R\x S^5} P[C^{(2)}]\wedge F\wedge F\ , 
\label{FFcoupling}
\end{equation}
can be integrated directly over the $CP^2$ directions, yielding 
\begin{equation}
S_{CS}= n^2 T_1\int_{\R \x S^1} P[C^{(2)}]\ ,
\label{FFcoupling2}
\end{equation}
where we have used again that $T_1 = 4\pi^2 T_5$. Even though in $AdS_ 5 \x S^5$ 
$C^{(2)}$ is zero,
 this coupling indicates that the magnetic flux is
 inducing $n^2$ D-string charge in the configuration. These strings are
wound around the fibre direction  $\chi$. 

Note that $n$ D3-brane charge is also induced in the configuration through the Chern-Simons coupling
\begin{equation}
S_{CS}=-T_5 \int_{ \R\times S^5} P[C^{(4)}]\wedge F\, ,
\end{equation}
with the D3-branes wrapped on the non-trivial two-cycles of the $CP^2$. However, only the charge 
at the intersection of the two D3-branes
contributes to the energy. In fact, expression (\ref{ED5}) is precisely of the form of a 
threshold BPS intersection for D1- and D5-branes, being the total energy just the sum of the
energies of each of the constituents.

Let us now discuss the influence of the $N$ fundamental strings
that stretch from the D5-brane to the boundary of $AdS_5$. 
In order to keep the spherical D5-brane undeformed the F-strings must be uniformly scattered over 
the five-sphere. Otherwise, if a significant number of strings are joined at the same point, their
backreaction is not negligible and they will start to deform the 5-sphere 
\cite{Imamura1,Imamura2}. Then, in this limit, one would need to consider the full DBI problem,
in which the F-strings are seen as a spike in the worldvolume of the D5-brane, in the
spirit of \cite{CGS, CGMvP, GRST}. However,
taking the $N$ fundamental strings to join the D5-brane in different points 
breaks all the supersymmetry, since although each string preserves 
one half, the fact that they take different positions on the D5-brane makes that
 the preserved Killing spinors of each string are different, such that all
supersymmetries are broken \cite{Imamura2}.  We will see below 
that this breaking results in the fact that the configuration has a binding energy.

The spherical D5-brane with magnetic flux that we have  discussed in this section is very similar to the 
spherical D2-brane probe with dissolved D0-brane charge of \cite{Emparan}. It was shown in
\cite{Myers} that there exists a complementary, microscopical description of this system 
in terms of D0-branes expanding by dielectric effect into a fuzzy spherical D2-brane, and that 
when the number of D0-branes is large enough, the microscopical and D2-brane descriptions
coincide. The analogy with our case suggests that there should
exist a microscopical description of the baryon vertex with magnetic flux, in terms of multiple
non-Abelian D-strings, expanding into a fuzzy spherical D5-brane. We will provide this 
microscopical description in the next section. In the remaining part of this section  
we will first analyse the influence of the magnetic 
flux on the dynamics of the baryon vertex. 

\subsection{The bound on the instanton number}

It was argued in \cite{BISY} that in order to analyse the stability of the baryon vertex
in the $u$-direction (i.e. against perturbations in the holographic direction of  $AdS$), 
one has to consider the influence of the external F-strings. 
The energy $E$ 
of the baryon vertex is then proportional to $N$ times the energy of a $q\bar q$ system, which is 
in turn inversely proportional to the distance $\ell$ between the quarks \cite{Maldacena2}.
As the proportionality constant between $E$ and $\ell$ is negative, the baryon vertex is 
indeed stable under perturbations in $u$. 

In this subsection we will perform the same calculation in \cite{BISY}, but taking into account 
the effect of the non-zero magnetic flux on the D5-brane. 

The action for the baryon vertex  
with magnetic flux on the worldvolume of the D5-brane is given by
\begin{equation}
S=S_{D5}+S_{NF1},
\end{equation}
with $S_{D5}$ given by minus the time integration of (\ref{ED5}). On the other hand, the F-strings 
connecting the D5-brane and a quark on the boundary can be parametrised by the worldvolume 
coordinates $\{t,x\}$ and the position in $AdS$ by $u=u(x)$. Then, the 
Nambu-Goto action is given by
\begin{equation}
S_{NF1}=-NT_1 \int dt dx \sqrt{(u')^2+\frac{u^4}{L^4}},
\end{equation}
where $u'$ denotes the derivative of $u(x)$ with respect to $x$.
Following the analysis of \cite{BISY}, the equations of motion
associated to
the system come in two sets: the bulk equation of motion
for the strings, and the boundary equation of motion (as we are 
dealing with open strings), which contains as well a term coming from the D5-brane.
One can show easily that these equations of motion are:
\bea
&& \frac{u^4}{\sqrt{(u')^2+\frac{u^4}{L^4}}}\ = \ {\rm const}, 
\label{bulk}\\ [.2cm]
&& \frac{u'_0}{\sqrt{(u'_0)^2+\frac{u_0^4}{L^4}}}
           \ = \ \frac{\pi L^4}{4N}\Bigl(1+\frac{8n^2}{L^4}\Bigr),
\label{boundary}
\eea
for the bulk and the boundary respectively, with $u_0$ the position of the baryon vertex in the 
holographic direction and  $u'_0 = u'(u_0)$. For future convenience, let us call
\begin{equation}
\sqrt{1-\beta^2}\ = \ \frac{\pi  L^4}{4N}\Bigl(1+\frac{8n^2}{L^4}\Bigr).
\label{T}
\end{equation}
Notice that in our conventions, $g_s=1$ and $2\pi l_s^2 =1$, we have that 
\be
L^4 \ = \ 4\pi g_sl_s^4N \ = \ \frac{N}{\pi}\, ,
\label{L(N)}
\ee
and we can rewrite (\ref{T}) as 
\begin{equation}
\beta^2\ = \ 1 - \frac{1}{16}\Bigl( 1 +\frac{8\pi n^2}{N}\Bigr)^2.
\label{beta}
\end{equation}
Equations (\ref{bulk}) and (\ref{boundary}) can then be combined into a single one,
\begin{equation}
\frac{u^4}{\sqrt{(u')^2+\frac{u^4}{L^4}}}\ =\ \beta \ u_0^2L^2.
\end{equation}
In the absence of magnetic BI flux on the worldvolume, $\beta =
\sqrt{15/16}$, as in 
\cite{BISY}. However, in general for non-zero $n^2$, 
we have to make sure that $\beta$ is real (as $u$ is real), which
from (\ref{beta}) 
implies that
\begin{equation}
\label{bound}
\frac{n^2}{N}\le\frac{3}{8\pi}.
\end{equation}
Surprisingly, we find that there is a bound on the number of D-strings
that can be dissolved in the configuration,
which depends on the number of D3-branes
that source the background. 

\begin{figure}[t]
\begin{center}
\label{radio}
\includegraphics{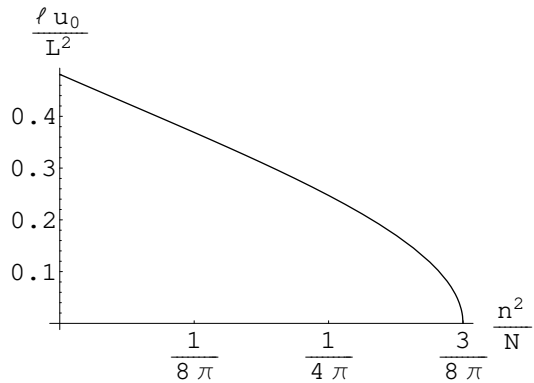}
\caption{\it Radius of the baryon (in units of $L^2/u_0$) as a function of $\frac{n^2}{N}$.}  
\end{center}
\end{figure}

Integrating the equation of motion, we find that the size of the
baryon $\ell$ is given by
\begin{equation}
\ell=\frac{L^2}{u_0}\int_1^{\infty} dy\frac{\beta}{y^2\sqrt{y^4-\beta^2}}, 
\end{equation}
with $y= u/u_0$.
This integral can be solved in terms of hypergeometric functions \cite{BISY}. In Figure 
1 we have plotted the radius $\ell$ of the baryon as a function of  
$\frac{n^2}{N}$. The plot reveals that the radius of the baryon cannot be continued
outside the allowed domain given by (\ref{bound}). Note that the size of the baryon vertex 
goes to zero  as we saturate the bound.

\begin{figure}[t]
\begin{center}
\includegraphics{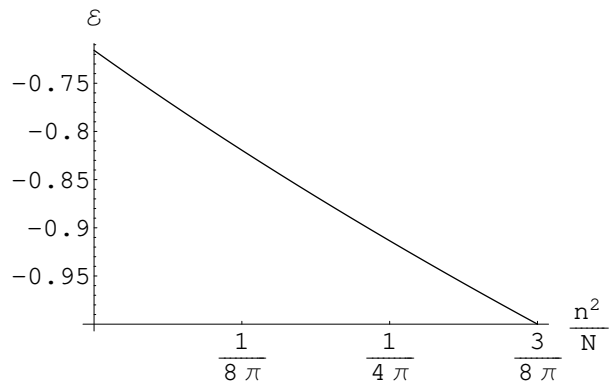}
\label{koekoek}
\caption{\it The energy (in units of $u_0$) of the baryon vertex as a function
 of $\frac{n^2}{N}$.}  
\end{center}
\end{figure}

Finally, the energy for a single string is, in terms of $u_0$:
\begin{equation}
E=T_1 u_0 \Bigr\{
    \int_1^{\infty}dy \Bigl[\frac{y^2}{\sqrt{y^4-\beta^2}}-1 \Bigr] 
      - 1 \ \Bigr\}.
\end{equation}
Notice that this expression has the same form than the expression in \cite{BISY}, and indeed
takes the same value for $n= 0$. In particular, the dependence on 
$\sqrt{g^2N}$ and on $u_0$ is unaltered, as expected by conformal invariance.
The explicit dependence of the energy on the ratio $\frac{n^2}{N}$ can be seen in 
Figure 2. As expected, the configuration is only well defined for $\frac{n^2}{N}$
inside the allowed interval. 

The fact that we find a bound on the number of dissolved D1-branes due to the dynamics of the 
F-strings, is quite surprising, and it is not entirely clear to us what its interpretation is. For 
a brief discussion we refer to the conclusions.

\sect{The microscopic description of the baryon vertex}

As we mentioned at the end of subsection 2.2, the fact that the magnetic flux in the D5-brane 
worldvolume induces D1-brane charge, suggests a close analogy with 
the dielectric effect described in \cite{Emparan, Myers}. In this section we will show that 
it is indeed possible to give an alternative, microscopic description of the baryon vertex, 
in terms of a fuzzy spherical D5-brane built up out of dielectrically expanded D1-branes. 

\subsection{D1-branes polarising to a 5-brane}

The action describing the dynamics of $n^2$ coinciding D1-branes is the
non-Abelian action given in \cite{Myers}, which for the $AdS_5 \x S^5$
background reduces to the form
\bea
\label{nD1}
S_{n^2D1} &=& - T_1 \int d^2\xi \ \STr \Bigl\{
       \sqrt{\Bigl|{\rm det}\Bigl(P[g_\mn + g_{\mu i}(Q^{-1} -\delta)^i{}_j 
                             g^{jk}g_{k\nu}]\Bigr) 
                                 \det Q\Bigr| \ }\Bigr\} \nnw
        && + \ T_1 \int d^2\xi \ \STr \Bigl\{
               P[ i (\incl_X \incl_X) C^{(4)} \ - \ \half (\incl_X \incl_X)^2
		 C^{(4)} \wedge {\cal F}\Bigr\},
\eea
where $g_\mn$ is the metric in  $AdS_5 \x S^5$ and 
\bea
&&Q^i{}_j = \delta^i_j + i [X^i, X^k] g_{kj}, \nn
&&\Bigl((\incl_X \incl_X) C^{(4)}\Bigr)_\mn 
         = \ \tfrac{1}{2}[X^\lambda, X^\rho] C^{(4)}_{\rho\lambda\mn}, \\
&&(\incl_X \incl_X)^2 C^{(4)} = \ \tfrac{1}{4}[X^\lambda, X^\rho ][X^\nu, X^\mu] C^{(4)}_{\mnrl}. 
	 \nonumber
\eea
Inspired by the coupling (\ref{FFcoupling2}) in the D5-brane calculation
we wind the D-strings around the $U(1)$ fibre direction $\chi$ and let
them expand into the $CP^2$. In this way we obtain a fuzzy version of the
$S^5$ as an Abelian $U(1)$ fibre over a fuzzy $CP^ 2$, similar to the
microscopic description of 5-dimensional giant gravitons in $AdS_4 \x S^7$ and
$AdS_7 \x S^4$ found in \cite{JLR3}. 

In the $AdS_5 \x S^5$ background the Chern-Simons couplings in (\ref{nD1})
vanish. Therefore, the expansion of the strings into a fuzzy $CP^2$ is caused by the
couplings in the Born-Infeld part of the action and, thus,
it is entirely due to 
a gravitational dielectric
effect, analogous to the configurations described in \cite{Diego}. 

A fuzzy version of $CP^2$ is well known (see for example \cite{ABIY}). $CP^2$
is the coset manifold $SU(3)/U(2)$, and can be defined as the submanifold of
$\R^8$ determined by the constraints
\be
\label{condi}
\sum_{i=1}^8 x^i x^i=1\ , 
\hspace{2cm}
\sum_{j,k=1}^8d^{ijk}x^j x^k =\frac{1}{\sqrt{3}}x^i\ ,
\ee
where $d^{ijk}$ are the components of the totally symmetric
$SU(3)$-invariant tensor. In our coordinates (\ref{CP2}) we have
(see \cite{JLR3})
\be
\begin{array}{ll}
x^1=\frac{\sqrt{3}}{2} \sin{2\varphi_1}\cos{\frac{\varphi_2}{2}}
         \cos{\frac{\varphi_4+\varphi_3}{2}},  \hspace{2cm} 
& x^5=-\frac{\sqrt{3}}{2} \sin{2\varphi_1}\sin{\frac{\varphi_2}{2}}
            \sin{\frac{\varphi_4-\varphi_3}{2}}, \\[.3cm]
x^2=-\frac{\sqrt{3}}{2} \sin{2\varphi_1}\cos{\frac{\varphi_2}{2}}
      \sin{\frac{\varphi_4+\varphi_3}{2}}, 
&x^6=\frac{\sqrt{3}}{2} \sin^2{\varphi_1}\sin{\varphi_2}\cos{\varphi_3}, \\[.3cm]
x^3=\frac{\sqrt{3}}{2}[\sin^2{\varphi_1}(1+\cos^2{\frac{\varphi_2}{2}})-1],
& x^7=-\frac{\sqrt{3}}{2} \sin^2{\varphi_1}\sin{\varphi_2}\sin{\varphi_3}, \\[.3cm]
x^4=\frac{\sqrt{3}}{2} \sin{2\varphi_1}\sin{\frac{\varphi_2}{2}}
      \cos{\frac{\varphi_4-\varphi_3}{2}},
& x^8=\frac12 (3\sin^2{\varphi_1}\sin^2{\frac{\varphi_2}{2}}-1)\, ,
\end{array}
\ee
for which $\frac13 \sum_{i=1}^8 (dx^i)^2 = ds^2_{CP^2}$ is the Fubini-Study metric (\ref{CP2}).
A  fuzzy version of $CP^2$ can  be obtained by
imposing the conditions (\ref{condi}) at the level of matrices. Define a
set of coordinates $X^i$ $(i=1,\dots,8)$ as
\begin{equation}
\label{defX}
X^i=\frac{T^i}{\sqrt{(2n^2-2)/3}},
\label{X(T)}
\end{equation}
with $T^i$ the generators of $SU(3)$ in the $n^2$-dimensional
irreducible representations $(k,0)$ or $(0,k)$, with 
$n^2=(k+1)(k+2)/2$ (see \cite{JLR3,ABIY} for more details). Nothe that
$(2n^2-2)/3$ is the quadratic Casimir of $SU(3)$ in
these representations. The first constraint in (\ref{condi}) is then trivially
satisfied through the quadratic Casimir of the group, whereas the rest of the
constraints are satisfied for any $n^2$.  The commutation relations between the
$X^i$ are given by
\be
[X^i, X^j] = \frac{i f^{ijk}}{\sqrt{(2n^2-2)/3}} X^k,
\label{[XX]}
\ee
with $f^{ijk}$ the structure constant of $SU(3)$ in the algebra of
the Gell-Mann matrices $[\lambda^i,\lambda^j]=2i f^{ijk} \lambda^k$.

Substituting the non-commutative Ansatz above in the action (\ref{nD1}) and
particularising to the $AdS_5\times S^5$ background, we find
\bea
S_{n^2D1} &=& - T_1 \int dt d\chi \ u \ 
               \Str\Bigl\{ \unity +\frac{L^4}{4(2n^2-2)}\unity\Bigr\}  \nnw
       &=& -2\pi n^2  T_1  \ \int dt \ u 
                           \ \Bigl( 1 +  \frac{L^4}{8(n^2-1)}\Bigr)\, ,
\eea
since, remarkably, 
\begin{equation}
{\rm det}Q=\Bigl(\unity+\frac{L^4}{4(2n^2-2)}\unity\Bigr)^2,
\end{equation} 
in the large $n$ limit\footnote{To be more precise, we are neglecting higher order powers of $L^2/n$. This is the right limit to study the matching with the macroscopical description of the previous section.}.
It is to be emphasized that the fact that the $\det Q$ is a 
perfect square is the microscopical analogous of the perfect square that we obtained for the DBI action in the 
macroscopic case. 

The energy of the $n^2$ expanded D1-branes is then given by
\be
E_{n^2D1} =  2\pi u  T_1 \ \Bigl( n^2 + \frac{n^2L^4}{8(n^2-1)}\Bigr). 
\label{EnD1}
\ee
Taking into account that the tensions of the D1- and the D5-brane are related by
$T_1 = 4\pi^2 T_5$, it is easy to see that in the limit where the number of D1-branes 
$n^2\rightarrow \infty$, the above expression reduces to the energy
of the macroscopic D5-brane, given by (\ref{ED5}).

\subsection{The $N$ F-strings in the microscopic description}

So far we have compared the energy of the spherical D5-brane of the
baryon vertex to the energy of the configuration built up by $n^2$ D1-branes
expanding into a D5-brane with the topology of a fuzzy 5-sphere.  We have shown 
that the two
descriptions agree in the limit where the instanton number on the D5-brane is
very large. However an essential part in the construction of the baryon vertex
are the $N$ fundamental strings that stretch from the D5-brane in the interior 
to the boundary of $AdS_5$. In this subsection we show how these strings arise 
in the microscopical setup.

The CS action for coincident D-strings contains the following couplings to the
$C^{(4)}$ R-R potential:
\bea
S_{CS}= T_1 \int dtd\chi \ {\rm STr} \Bigl\{ P[(\incl_X \incl_X)C^{(4)}] 
             \ - \  P[(\incl_X \incl_X)^2C^{(4)}]\wedge {\cal F}\  \Bigr\}\, , 
\label{nonAbCS}
\eea
where ${\cal F} = d{\cal A} + [{\cal A}, {\cal A}]$ is the $U(n^2)$ BI field strength.

The first term in (\ref{nonAbCS}) is zero in the $AdS_5 \x S^5$ background.
The second term, in turn, can be written as
\begin{equation}
\label{CSD1}
S_{CS}= \frac{T_ 1}{4} \int dtd\chi \ {\rm STr}\Bigl\{[X^i,X^j][X^k,X^l]C^{(4)}_{ijkl} 
               \ \partial_{\chi} {\cal A}_t \Bigr\},
\end{equation}
in the gauge ${\cal A}_\chi = 0$.  Integrating by parts we have that
\begin{equation}
S_{CS}=-\frac{T_1}{4}\int dt d\chi \ {\rm STr}\Bigl\{[X^i,X^j][X^k,X^l]
G^{(5)}_{\chi ijkl}{\cal A}_t\Bigr\}.
\end{equation}
Taking into account that in the non-commutative coordinates introduced in
(\ref{X(T)}), $G^{(5)}$ is given by \cite{JLR3}
\begin{equation}
G^{(5)}_{\chi ijkl}=L^4 f_{[ij}^m f_{kl]}^n X^m X^n,
\end{equation}
we find that
\be
S_{CS} =  \frac{L^4 T_ 1}{2(n^2-1)} \int dtd\chi\ {\rm STr} \Bigl\{{\cal A}_t \Bigl\},
\ee
where we have made use of the commutation relations (\ref{[XX]}). 
In analogy with the Abelian case (\ref{AnsatzA}), we can take as an Ansatz for ${\cal A}$,
\begin{equation}
{\cal A}=A_t (t) \unity dt.
\end{equation}
Integrating over $\chi$ and taking into account that $L^4=N/\pi$ we find
finally that
\be
S_{CS} =  \frac{n^2}{n^2-1}\ N T_1 \int dt \  A_t.   
\label{SCSD1FULL}
\ee
The coupling (\ref{CSD1}) is therefore inducing, in the large $n^2$ 
   limit, $N$ BI charges in the configuration.\footnote{The fact that the number of strings 
   is $N$ only in the large $n^2$ limit is similar to the construction of the fuzzy funnels of 
   \cite{CMT, CMT2}, where D3- and D5-branes are shown to have integer charges only in the limit 
   of infinite D1-branes.} These charges have to be cancelled by $N$ fundamental strings ending on the
D1-brane system. The dielectric coupling to $C^{(4)}$ in
(\ref{nonAbCS}) will then take care that 
these strings are expanded over the full $S^5$.

We can therefore conclude that our microscopical picture, consisting of multiple coinciding 
D-strings expanding into a fuzzy D5-brane, reproduces in the large $n^2$ limit all the relevant 
features of the baryon vertex with magnetic flux. Not only did we obtain the same expression for 
the energy of the D5-brane, but we also found traces of the presence of the fundamental strings 
stretched between the (dielectric) 5-brane and the boundary of $AdS$. 

\subsection{F-strings polarising to a NS5-brane}

Due to the S-duality invariance of the $AdS_5\times S^5$ background, the baryon
vertex with magnetic flux, described in the previous sections, can alternatively be 
realised as a NS5-brane wrapped on the $S^5$, with
$N$ D1-branes stretching between the brane and the boundary of the $AdS$ space and
with $n^2$ F-string charge dissolved in its worldvolume.
Microscopically this configuration is described in terms of
fundamental strings expanding into a fuzzy NS5-brane. In this subsection we give the
details of this description.

An action describing coincident F-strings in Type IIB can be constructed from the action
for coincident gravitational waves in Type IIA, using T-duality. Such an action was 
constructed in \cite{BJL} to the linearised level in the background fields, and 
turned out to be the S-dual of the action for coincident D1-branes of \cite{Myers},
linearised in the background fields. Since in this picture the dynamics of the non-Abelian 
F-strings is induced by the open D-strings that end on them, this action is adequate to 
describe the system in the strong coupling regime. 

However, given the S-duality invariance of the 
$AdS_5\times S^5$ background, the non-Abelian action for F-strings can well
be used here. We will start by constructing
an action valid beyond the linearised
level, and therefore suitable for the study of the $AdS_5 \times S^5$ background.

Using the action of
\cite{JL1,JL2} for coincident Type IIA gravitational waves,
valid to all orders in the background fields, we can
construct an action describing coincident Type IIB F-strings by T-dualising along
the direction of propagation of the waves.

The action for coincident Type IIA gravitational waves contains a worldvolume scalar field
associated to D0-branes ``ending" on the system (see \cite{JL1}). We will set to zero this field
for simplicity and take as well $B^{(2)} = C^{(1)} = 0$.  This is suitable for the study
of the $AdS_5\times S^5$ background. We then have (see \cite{JL1,JL2})
\bea
S_{n^2W_A} &=& -T_W\int d\tau\ \Str\Bigl\{ 
   k^{-1} \sqrt{\Bigl| \det\Bigl(P[E_{\mn}+E_{\mu i}(Q^{-1}-\delta)^i{}_j 
          E^{jk}E_{k\nu}] \Bigr) 
                     \ {\rm det} Q  \Bigr|\  } \Bigr\} \nnw
  && \hspace{-1cm}
+ \ T_W \int d\tau\ \STr \Bigl\{
-  P [k^{-1} k^{(1)} ] +  i   P [(\incl_X \incl_X)C^{(3)}] 
                +\frac{1}{2} P[(\incl_X \incl_X)^2 \incl_k B^{(6)}]
 \Bigr\},
\label{waveaction}
\eea
where
\begin{eqnarray}
E_{\mu\nu} &=& g_{\mu\nu}-k^{-2}k_\mu k_\nu
                   + k^{-1}e^\phi (i_k C^{(3)})_{\mu\nu},  \nnw
Q^i{}_j &=& \delta^i_j \ + \ i e^{-\phi}k[X^i ,X^k] E_{kj}\, .
\end{eqnarray}
Here $k^\mu$ is a Killing vector pointing on the direction of propagation of the gravitational 
waves.\footnote{In our notation $k^{(1)}=g_{\mu\nu}k^\nu dx^\mu$. The
  coupling to $k^{(1)}$ in (\ref{waveaction}) shows
that the waves carry momentum along the Killing direction. See \cite{JL1} for more
details.} $B^{(6)}$ is the NS-NS 6-form potential. Note that
(\ref{waveaction}) is a gauged sigma model, in which the Killing direction
does not appear as a physical degree of freedom \cite{BJO}.

T-dualising the above action along the Killing direction, we get a non-Abelian
action for $n^2$ F-strings in Type IIB:
\begin{eqnarray}
S_{n^2F1} &=&-T_1\int d\tau d\sigma\ \Str\Big\{
\sqrt{\Bigl|{\rm det}\Bigl(P[E_\mn + E_{\mu i}(Q^{-1}-\delta)^i{}_j 
                    E^{jk}E_{k\nu}]\Bigr)
           \det Q \Bigr|}\ \Big\}\nnw 
&& \hspace{.5cm}      -T_1\int d\tau d\sigma \ \Str\Big\{P[B^{(2)}]  
      + \ i  P[(\incl_X\incl_X) C^{(4)}]
      -\frac12 P[(\incl_X \incl_X)^2 B^{(6)}] \Big\},
\label{F1action}
\end{eqnarray}
where now
\bea
E_{\mu\nu} &=& g_{\mu\nu} + e^\phi C^{(2)}_{\mu\nu},  \nnw
Q^i{}_j &=& \delta^i_j \ +\ ie^{-\phi}[X^i,X^k] E_{kj}.
\eea
This action is no longer a gauged sigma model, as it can
be written in a completely covariant way.  Although some of the fields are set to
zero due to the truncation in (\ref{waveaction}) we see that (\ref{F1action}) is just
the S-dual of the action for $n^2$ coincident D-strings of \cite{Myers}. In particular, the
non-Abelian
worldvolume scalar associated to D0-branes ending on the Type IIA waves is mapped under 
T-duality into a non-Abelian vector field which is now associated to D1-branes ending on the
system of fundamental strings. One can check at the linearised level (see \cite{BJL})
that the field strength of this vector field appears
in the action for the F-strings exactly as predicted by S-duality.

Using the action (\ref{F1action})
to describe $n^2$ F-strings in the $AdS_5\times S^5$ background is now
straightforward. The computation of the energy of the baryon vertex
reduces to the same computation of subsection 3.1. However
the strings expand now into a fuzzy NS5-brane, since the configuration acts as a source for 
the $B^{(6)}$ potential through the last
coupling in (\ref{F1action}).  
 The energy of the configuration is given by
\be
E_{n^2 F1} = 2\pi u  T_1 \ \Bigl( n^2 + \frac{n^2L^4}{8(n^2-1)}\Bigr), 
\ee
which matches exactly the result (\ref{EnD1}) obtained from
the D1-brane calculation. 

Finally, the S-dual of the coupling (\ref{CSD1})  shows that $N$ open D1-branes must
be added to the configuration stretching between the NS5-brane and the boundary
of the $AdS$ space. 
Therefore we have provided a microscopical description of the (generalised) baryon
vertex in terms of a spherical NS5-brane with $N$ D1-branes attached  to it \cite{Witten}.

\sect{Conclusions}

The baryon vertex  consists on a single probe 
D5-brane wrapping the $S^5$ in $AdS_5\times S^5$ to which $N$ fundamental
strings are attached, running from the D5-brane to the $AdS$ boundary. Since all the
strings have the same orientation, this represents a gauge invariant bound state 
of $N$ quarks, \textit{i.e.} a baryon \cite{Witten}. Due to the S-duality invariance of the
$AdS_5\times S^5$ background the baryon vertex can alternatively be realised in terms
of a NS5-brane with $N$ D1-branes attached or as a $(p,q)$ 5-brane with $(p,q)$ strings
attached. 

In this letter, we have found a generalised version of the baryon vertex by writing
the $S^5$ as an $S^1$ fibre bundle over $CP^2$. Since $CP^2$ admits an instantonic 
magnetic field proportional to the curvature tensor of the fibre connection, it is possible to
consistently plug in a magnetic field on the worldvolume of  the D5-brane.
These instantons have the effect of dissolving a number $n^2$ of D-strings on the D5-brane, 
wound in the  fibre direction. This charge is, however, not
topological, since the 5-sphere has no non-trivial cycles. Therefore,
it might be that the baryon vertex with flux represents only a
metastable configuration\footnote{We thank the referee for a useful discussion
  about this point.} (see fro example \cite{KPV, ISS, ABFK} for a recent discussion on this type of configurations).

On the other hand,
the fact that one can consistently add a number of dissolved D1-branes to the worldvolume of 
the D5, hints to the existence of an alternative description of the baryon vertex, in terms of 
expanded D1-branes.  We have explicitly provided such a microscopical description of the 
generalised baryon vertex in terms of D-strings (F-strings) expanding into a D5-brane (NS5-brane) 
due to Myers dielectric effect. Here the dielectric effect is purely gravitational, i.e. caused 
by the curvature of the background. Indeed, the CS coupling, as in the macroscopical case, is 
only indicating the need to introduce the $N$ external F-strings (D-strings) that build up the
vertex. The expanding strings are wound along the $S^1$ fibre of $S^5$ and expand into
a fuzzy version of $CP^2$. The fuzzy $S^5$ is then realised as an Abelian $U(1)$ fibre
over a fuzzy $CP^2$.  

Our construction needs, implicitly, that the F-strings are
uniformly scattered over the D5-brane, in such a way that their backreactions are
compensated and the D5-brane remains approximately spherical. This however has
the effect of breaking all the supersymmetries \cite{Imamura2}. If we insist in preserving some
of the supersymmetries, we have to let all strings end at the same point of the D5-brane, 
which in turn invalidates our probe approximation. Indeed, in such a case one should look for a full
description of the baryonic brane \cite{Imamura1,Imamura2}, in terms of a single
D5-brane developing a spike representing the F-strings \cite{CGS}-\cite{GRST}, 
analogous to the D3-brane spike of \cite{callan}. However, while in that case the
binding energy of the configuration is zero, reflecting the fact that it is supersymmetric, 
this is not the case for our configuration, for which we obtain a non-zero binding
energy. 

One of the most surprising conclusions of the analysis of the dynamics of the generalised baryon 
vertex is the fact that the number of dissolved D-strings is bounded from above. A careful study of 
the baryon vertex along the lines in \cite{BISY}, shows that the configuration is stable against
fluctuations in the $u$ direction. In particular, the configuration has the same dependence of
the energy on $u_0$ as the original vertex. However, in our case the number of dissolved strings 
must not violate certain bounds imposed by the dynamics of the system. It is likely that this bound 
is related to the stringy exclusion principle of \cite{MaldaStrom}. Our configuration with non-zero
magnetic flux carries a non-zero winding number in the fibre direction of the $S^5$, which in 
terms of the dual field theory will manifest itself as a charge under a specific $U(1)$ subgroup of 
the $SU(3)$ R-symmetry group \cite{BHK}. As these charges are bounded due to conformal
invariance,
one expects to find a bound on the magnetic flux. This is quite similar to the giant graviton effect.
Indeed, in $AdS_5\times S^5$ there exists a giant graviton, which consists of a 
D3-brane wrapped on an $S^3$ inside the $CP^2$ part of the $S^5$ and moving along the
fibre direction. This giant graviton state corresponds in the dual field theory to a chiral 
primary operator with the same $U(1)$-charge as our configuration. It is surprising however that we 
can only find the bound on the magnetic flux when the whole system of wrapped D5-branes and 
F-strings is considered. Another interesting observation is made in
\cite{SJ}, where it was noted that in $AdS_p \times S^q$ spaces the
radius $L$ of the $S^q$ can be expressed in terms of the dimension $n$
of the representation as $L^{q-1}=l_{PL}^{q-1} n$, if one tries to
describe this $q$-sphere as a fuzzy manifold.
We leave the precise interpretation
of this charge to future investigations.

\vspace{.4cm}
\noindent
{\bf Acknowledgements}\\
We wish to thank J.L.F. Barb\'on, C. Hoyos, J. Hirn, P. Meessen, S. Montero, T. \Ortin, 
J.P. Resco,  V. Sanz and especially C. Nu\~nez and A.V. Ramallo for useful discussions. 
The work of B.J. is done as part of the program ``Ram\'on y Cajal'' of the M.E.C. (Spain). 
He was also partially supported by the M.E.C. under contract FIS 2004-06823 and  by the 
Junta de Andaluc\'{\i}a group FQM 101.  
The work of Y.L. has been partially supported by CICYT grant BFM2003-00313 (Spain) and by 
the European Commission FP6 program MRTN-CT-2004-005104, in which she is associated to Universidad
Aut\'onoma de Madrid.
D.R.G. would like to thank the Departamento de F\'{\i}sica Te\'orica y del Cosmos of the
Universidad de Granada for its hospitality while part of this work was done, and for its
financial support through a visiting professor grant. He is also grateful
to the Departamento de F\'{\i}sica de Part\'{\i}culas of the Universitad de
Santiago de Compostela for their hospitality in the final stages of this work.




\begin{thebibliography}{99}

\bibitem{Maldacena} J. Maldacena, Adv. Theor. Math. Phys. 2 (1998) 231, hep-th/9711200.

\bibitem{ESZ} J. Erickson, G. Semenoff, K. Zarembo,
              Nucl. Phys. B582 (2000) 155, hep-th/0003055.

\bibitem{Maldacena2} J. Maldacena, Phys. Rev. Lett. 80 (1998) 4859, hep-th/9803002.

\bibitem{RY} S. Rey, J. Yee, Eur. Phys. J. C22 (2001) 379, hep-th/9803001.

\bibitem{Witten} E. Witten, JHEP 9807 (1998) 006, hep-th/9805112. 

\bibitem{MaldaStrom} J. Maldacena, A. Strominger, JHEP 9812 (1998) 005, hep-th/9804085. 

\bibitem{BHK} D. Berenstein, C.P. Herzog, I.R. Klebanov, JHEP 0206
  (2002) 047, hep-th/0202150.

\bibitem{Myers} R. Myers,  JHEP 9912 (1999) 022, hep-th/9910053.

\bibitem{JL2} B. Janssen, Y. Lozano, Nucl. Phys. B658 (2003) 281, hep-th/0207199.

\bibitem{BISY} A. Brandhuber, N. Itzhaki, J. Sonnenschein, S. Yankielowicz, 
              JHEP 9807 (1998) 020, hep-th/9806158.

\bibitem{Trautman} A. Trautman, Int. J. Theor. Phys. 16 (1977) 561.

\bibitem{HP} S.W. Hawking, C.N. Pope, Phys. Lett. B73 (1978) 42.

\bibitem{Pope} C.N. Pope, Phys. Lett. B97 (1980) 417.

\bibitem{Imamura1} Y. Imamura, Prog. Theor. Phys. 100 (1998) 1263  hep-th/9806162.

\bibitem{Imamura2} Y. Imamura, Nucl. Phys. B537 (1999) 184, hep-th/9807179.

\bibitem{CGS} C. Callan, A. Guijosa, K. Savvidy, Nucl. Phys. B547 (1999) 127, hep-th/9810092.

\bibitem{CGMvP} B. Craps, J. Gomis, D. Mateos, A. Van Proeyen, JHEP 9904 (1999) 004, hep-th/9901060.

\bibitem{GRST} J. Gomis, A. Ramallo, J. Sim\'on, P. Townsend, JHEP 9911 (1999) 019, hep-th/9907022.


\bibitem{Emparan} R. Emparan, Phys. Lett. B423 (1998) 71, hep-th/9711106.

\bibitem{JLR3} B. Janssen, Y. Lozano, D. Rodr\'{\i}guez-G\'omez, 
               Nucl. Phys. B712 (2005) 371,  hep-th/0411181.

\bibitem{Diego}  D. Rodr\'{\i}guez-G\'omez, JHEP 0601 (2006) 079, hep-th/0509228.

\bibitem{ABIY} G. Alexanian, A.P. Balachandran, G. Immirzi, B. Ydri, 
               J. Geom. Phys. 42 (2002) 28, hep-th/0103023.

\bibitem{CMT} N. Constable, R. Myers, \O. Tafjord, Phys. Rev. D61 (2000) 106009, hep-th/9911136.

\bibitem{CMT2} N. Constable, R. Myers, \O. Tafjord, JHEP 0106 (2001) 023,
               hep-th/0102080.

\bibitem{BJL} D. Brecher, B. Janssen, Y. Lozano, Nucl. Phys. B634 (2002) 23, hep-th/0112180.

\bibitem{JL1} B. Janssen, Y. Lozano, Nucl. Phys. B643 (2002) 399, hep-th/0205254.

\bibitem{BJO} E. Bergshoeff, B. Janssen, T. Ort\'{\i}n,
              Phys. Lett. B410 (1997) 131, hep-th/9706117.

\bibitem{callan} C. Callan, J. Maldacena, Nucl. Phys. B513 (1998) 198, hep-th/9708147. 

\bibitem{KPV} S. Kachru, J. Pearson, H. Verlinde, JHEP 0206 (2002) 021, hep-th/0112197.

\bibitem{ISS} K. Intriligator, N. Seiberg, D. Shih, JHEP 0604 (2006) 021, hep-th/0602239.

\bibitem{ABFK} R. Argurio, M. Bertolini, S. Franco, S. Kachru, {\it Gauge/gravity duality and meta-stable dynamical supersymmetry breaking}, hep-th/9610212.

\bibitem{SJ} M.M. Sheikh-Jabbari, {\it Inherent Holography in Fuzzy
  Spaces and an N-tropic Approach to the Cosmological Constant
  Problem}, hep-th/0605110.





\end{thebibliography}
\end{document}